\documentclass[twocolumn,prb,superscriptaddress,floatfix,amsmath,amssymb,showpacs]{revtex4-1}
\usepackage{mathrsfs,bbm}%
\usepackage{graphicx}%
\usepackage{prettyref}%
	\newcommand{\pref}[1]{\prettyref{#1}}%
	\newrefformat{fig}{Fig.~\ref{#1}}%
	\newrefformat{tab}{Table~\ref{#1}}%
	\newrefformat{sec}{Sec.~\ref{#1}}%
	\newrefformat{eq}{Eq.~(\ref{#1})}%

\newcommand{\pwcond}{\texttt{PWcond}}
\newcommand{\pwscf}{\texttt{PWscf}}
\newcommand{\qe}{\textsc{Quantum~ESPRESSO}}
\newcommand{\etot}{\ensuremath{E_{\mathrm{tot}}}}%
\newcommand{\de}{\partial}
\newcommand{\Cdv}{\ensuremath{\mathrm{C_{2v}}}}
\newcommand{\Dqh}{\ensuremath{\mathrm{D_{4h}}}}
\newcommand{\ang}{\ensuremath{\textnormal{\AA}}}
\newcommand{\ev}{\ensuremath{{\mathrm{eV}}}}%
\newcommand{\ef}{\ensuremath{E_{\mathrm{F}}}}%
\newcommand{\echem}{\ensuremath{E_{\mathrm{chem}}}}%
\newcommand{\dco}{\ensuremath{d_{\mathrm{C-O}}}}%
\newcommand{\dauc}{\ensuremath{d_{\mathrm{Au-C}}}}%
\newcommand{\dwire}{\ensuremath{d_{\mathrm{chain}}}}%
\newcommand{\dauau}{\ensuremath{d_{\mathrm{Au-Au}}}}%
\newcommand{\nau}{\ensuremath{N_{\mathrm{Au}}}}%
\newcommand{\dsurf}{\ensuremath{d_{\mathrm{ss}}}}%
\newcommand{\dsm}{\ensuremath{d_{\mathrm{s1}}}}%
\newcommand{\dab}{\ensuremath{d_{1-2}}}%
\newcommand{\dbc}{\ensuremath{d_{2-3}}}%
\newcommand{\db}{\ensuremath{d_{\mathrm{b}}}}%
\newcommand{\ds}{\ensuremath{d_{\mathrm{s}}}}%
\newcommand{\kperp}{\ensuremath{\mathbf{k}_{\perp}}}%
\newcommand{\bfk}{\ensuremath{\textbf{k}}}%
\newcommand{\sga}{\ensuremath{5\sigma_{\mathrm{a}}}}%
\newcommand{\tps}{\ensuremath{2\pi^{\star}}}%
\begin{document}

\title{Effect of stretching on the ballistic conductance of Au nanocontacts in presence of CO: a density functional study
}
\date{\today}
\author{Gabriele Sclauzero}
\altaffiliation[Present address: ]{%
Ecole Polytechnique F\'ed\'erale de Lausanne (EPFL), ITP-CSEA, CH-1015 Lausanne, Switzerland}
\author{Andrea \surname{Dal Corso}}
\affiliation{%
International School for Advanced Studies (SISSA-ISAS), Via Bonomea 265, IT-34136 Trieste, Italy}
\affiliation{%
IOM-CNR Democritos, Via Bonomea 265, IT-34136 Trieste, Italy}
\author{Alexander Smogunov}
\altaffiliation[Present address: ]{%
CEA Saclay, IRAMIS/SPCSI, Bat. 462, 91191 Gif sur Yvette, France}
\affiliation{%
IOM-CNR Democritos, Via Bonomea 265, IT-34136 Trieste, Italy}
\affiliation{%
International Centre for Theoretical Physics (ICTP), Strada Costiera 11, IT-34151 Trieste, Italy}
\affiliation{%
Voronezh State University, University Sq. 1, 394006 Voronezh, Russia}
\begin{abstract}
CO adsorption on an Au monatomic chain is studied within density functional theory in nanocontact geometries as a function of the contact stretching.
We compare the bridge and atop adsorption sites of CO, finding that the bridge site is energetically favored at all strains studied here.
Atop adsorption gives rise to an almost complete suppression of the ballistic conductance of the nanocontact, while adsorption at the bridge site results in a conductance value close to $0.6\,G_0$, in agreement with previous experimental data.
We show that only the bridge site can qualitatively account for the evolution of the conductance as a function of the contact stretching observed in the experimental conductance traces.
The numerical discrepancy between the theoretical and experimental conductance slopes is rationalized through a simple model for the elastic response of the metallic leads.
We also verify that our conductance values are not affected by the specific choice of the nanocontact geometry by comparing two different atomistic models for the tips.
\end{abstract}

\maketitle

\section{Introduction}

As demonstrated by a recent experiment,\citep{kiguchi2007} the atomic conductance of a Au nanocontact can be altered by admitting gaseous CO in proximity of the nanocontact, suggesting that CO chemically interacts with the Au atoms in the thinnest part of the contact (a monatomic chain or a single atom). 
Indeed, besides a slight renormalization of the typical conductance peak at $1\,G_0$, the conductance histogram of Au after exposition to CO displays an additional peak between $0.5\,G_0$ and $0.6\,G_0$, which can be attributed to CO adsorption.
This is further supported by the analysis of the conductance traces, which clearly pinpoints the existence of an atomic structure with a CO molecule adsorbed on, or incorporated into the monatomic chain.
This structure is invariably characterized by a conductance value close to that of the new histogram peak and also by a small increase of the conductance upon further pulling of the nanocontact, just prior to contact breaking.\citep{kiguchi2007} 

In a previous density functional study, we addressed the adsorption of CO on Au monatomic chains using an infinite chain geometry without tips.\cite{sclauzero2012a}
The bridge adsorption site was found to be energetically favored with respect to the atop site, both at the equilibrium spacing of the Au chain and at larger values of the Au-Au spacing. 
We characterized the adsorption process by identifying the bonding/antibonding pairs of $5\sigma$ and \tps\ states which arise from the hybridization between CO molecular levels and Au metal states. 
Therefore, the electronic structure of this chain/adsorbate system results from a donation/backdonation mechanism analogous to the Blyholder model for CO on transition metal surfaces,\cite{blyholder1964,hammer1996,fohlisch2000} as previously demonstrated also for CO on Pt monatomic chains.\cite{sclauzero2008a,sclauzero2008b}

A strong connection between some electronic structure features of the CO/Au-chain system and the effects produced by CO adsorption on the ballistic transport across the chain was established in that work.\cite{sclauzero2012a}
In particular, the coupling of the $5\sigma$ antibonding state with the $s$ states of the Au chain was found to generate a dip in the $s$ transmission.
Hence, the position of this state turns out to be crucial in determining the reduction of the (tipless) conductance due to the impurity, as well as the strain dependence of the conductance in presence of CO. 
In the atop geometry, the $5\sigma$ antibonding state is located very close to the Fermi level (\ef) and approaches \ef\ as the Au chain gets stretched, hence it strongly reduces the conductance of a moderately strained chain and cuts almost completely the conductance of a highly strained chain (close to the rupture Au-Au spacing).\cite{calzolari2004,strange2008,sclauzero2012a}
In the bridge geometry instead, the conductance reduction is much lower than in the atop geometry because the $5\sigma$ antibonding state is more distant from \ef\ at medium strains and the associated transmission dip actually disappears at high strains.\cite{sclauzero2012a,sclauzero2012c}
At variance with the atop geometry, which shows a decrease of the conductance with strain and is therefore not compatible with the experimental conductance traces, the conductance of the bridge geometry shows the correct dependence on strain, namely, a slight increase with increasing strain.\cite{sclauzero2012a}

The calculated conductances of the tipless bridge geometry are however too large compared to the experimental conductance of the histogram peak appearing upon CO adsorption.\cite{kiguchi2007} 
Moreover, the nonuniform distribution of the strain, the finite length of the chain, and the effect of the tips are not described in the infinite chain model, therefore those results need to be corroborated by a more realistic modeling of the nanocontact.
Density functional calculations for various kinds of impurities adsorbed on model Au nanocontacts (for instance, molecular hydrogen\citep{csonka2003,csonka2006} and oxygen\citep{thijssen2006}), for CO on other metals, such as Pt,\cite{strange2006} and also for one specific geometry of CO on a Au nanowire contact\cite{xu2006} are already available in the literature.
However, different adsorption geometries of CO on Au nanocontacts have not been compared and the effect of contact stretching on the ballistic conductance of this system still needs to be analyzed.

\begin{figure}[tb]
	\begin{center}
		\includegraphics[width=0.9\columnwidth]{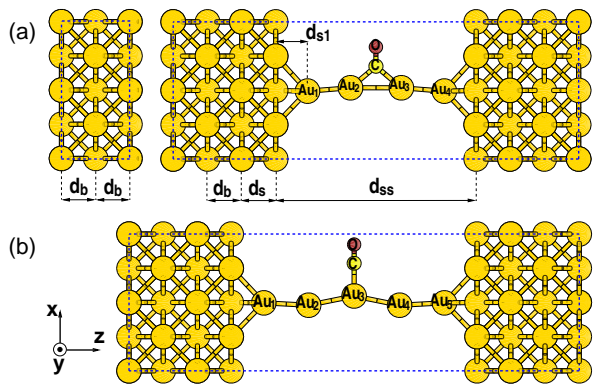}
	\end{center}
    \caption{(Color online) Lateral view of the periodic cells for the Au chain between Au(001) surfaces and CO adsorbed at the (a) bridge and (b) atop sites.
    The smaller cell in (a) is used to compute the complex band structure of the leads.
    The spacing between the bulk Au layers (\db), the outermost interlayer distance (\ds), and the inter-surface distance (\dsurf) are indicated.}
	\label{fig:geom}
\end{figure}

In this work, we study the adsorption energetics of CO on a short Au monatomic chain and we examine the ballistic conductance of the system as a function of the contact stretching using model nanocontact geometries.
By comparing the bridge and atop adsorption geometries of CO, we find that the bridge site is energetically favored at all strains, confirming the result obtained within the infinite chain model.\cite{sclauzero2012a} 
In the atop geometry, the Au conductance cut due to the $5\sigma$ antibonding states\cite{sclauzero2012a} is observed also in the present nanocontact model when the contact is sufficiently stretched. 
The bridge geometry instead, reproduces to a good accuracy the experimental conductance value ($0.5\div0.6\,G_0$),\cite{kiguchi2007} representing a significant improvement with respect to the infinite chain model.
The slope of the conductance as a function of the contact stretching is instead reproduced only qualitatively by our nanocontact model.
We address this discrepancy showing that a conductance slope closer to the experimental one can be obtained by taking into account, even only approximately, the effect of the elastic response of the bulk leads to the external pulling force, not described by our nanocontact model. 

This paper is organized as follows: in \pref{sec:method}, we describe the numerical methods and approximations adopted in the calculations; in \pref{sec:geom} and in \pref{sec:cond} we will report on, respectively, the adsorption energetics and the ballistic conductance of the system as a function of the contact stretching; \pref{sec:disc} is devoted to a discussion of the conductance results in the light of the available experimental data; finally, our conclusions will follow in \pref{sec:concl}.

\section{Methods and computational details}\label{sec:method}

All density functional calculations presented in this work are carried out using the plane-wave pseudopotential code \pwscf\ contained in the \qe\ package.\cite{QE-2009} 
The exchange and correlations functionals, both for the local density (LDA) and the generalized gradient approximation (GGA), the basis set cutoffs, the pseudopotentials, and the smearing parameters are the same as those used for the infinite chain model.\cite{sclauzero2012a}.

The monatomic chain in the nanocontact geometries is modeled as a row of metal atoms suspended between fcc bulk leads terminated by two facing (001) surfaces.\cite{smogunov2008b}
In \pref{fig:geom}, we show the simulation cells consisting of a periodically repeated slab geometry with 7 fcc-Au layers perpendicular to the [001] direction, which is chosen as $z$ axis and coincides with the electron transport direction (see later).
The spacing between the inner layers corresponds to the theoretical equilibrium value in the bulk \db, which is $2.029\,\ang$ ($2.082\,\ang$) according to our LDA (GGA) calculations.
The clean unreconstructed Au(001) surface shows a significant inward relaxation of about $1.8\%$ ($1.5\%$) for outermost layer within the LDA (GGA), in agreement with recent DFT calculations,\citep{singh-miller2009} while spacings between inner layers deviate from the bulk value by less than $0.4\%$.
Therefore, we keep into account the relaxation of the outermost layer on both sides of the junction by setting the first interlayer distance $\ds=1.992\,\ang$ ($2.051\,\ang$).
The in-plane distances between gold atoms are those of the bulk.
The short chain is freely suspended in the vacuum region between the slabs and the apex atoms of the chain are attached to the surfaces at a 4-fold coordinated hollow site. 
The symmetry group of the straight chain between the two (001) surfaces is \Dqh, the same as in the clean surface case studied with a slab geometry having inversion symmetry.

A single CO molecule is adsorbed at the center of the Au chain, at the bridge site of a 4-atom-long chain (\pref{fig:geom}a) and at the atop site of a 5-atom-long chain (\pref{fig:geom}b).
The symmetry group of the system in presence of CO is \Cdv, as in the case of CO adsorbed at the bridge or atop sites of the infinite chain.
The $xy$ in-plane periodicity of the simulation cells corresponds to a $(2\sqrt{2}\times2\sqrt{2})\mathrm{R}45^{\circ}$ surface structure, which gives a chain-chain spacing of about $8.12\,\ang$ in the $x$ and $y$ directions between two adjacent replicas.
The full Brillouin zone (BZ) of these structures is sampled with a uniform mesh of $6\times 6\times 3$ \bfk-points, which can be reduced by symmetry to 12 points (18 when CO is adsorbed).
We relax the atomic positions of the Au chain and of CO until the forces on those atoms drop below $0.026\;\ev/\ang$.
We also check that the optimized atomic positions of the chain and of CO obtained in this way result in atomic forces below $0.08\,\ev/\ang$ when used in a $(3\sqrt{2}\times3\sqrt{2})\mathrm{R}45^{\circ}$ cell (chain-chain spacing of $12.18\,\ang$ along $x$ and $y$).
For this larger cell we reduce the size of the \bfk-point mesh down to $4\times 4\times 3$.

The ballistic conductance is evaluated with the Landauer-B\"uttiker formula, $G = e^2/h\:T(E_F)$, where $T(E_F)$ is the total transmission at the Fermi energy.
We calculate the electron transmission using the scattering-based approach of \citeauthor{choi1999}\cite{choi1999} extended to ultrasoft pseudopotentials,\cite{smogunov2004b} and implemented in the \pwcond\ code.\cite{QE-2009}
In transmission calculations, we include in the scattering region a portion of the leads together with the impurity region (see \pref{fig:geom}).
The semi-infinite bulk leads are modeled using an additional, smaller cell which coincides with two bulk Au(001) layers (smaller cell in \pref{fig:geom}a).\footnote{The lead regions are used to compute the electronic complex band structures (CBSs) needed to solve the scattering problem. 
We verified that energy eigenvalues at real $k_z$ in the CBS obtained from the self-consistent potential in the leftmost part of the scattering region (corresponding to the unit cell of the lead) match within $0.05\,\ev$ the eigenvalues in the CBS of the lead region.}
In the $(2\sqrt{2}\times2\sqrt{2})\mathrm{R}45^{\circ}$ nanocontact geometry (which we will also call ``abrupt'' junction), the \kperp-dependent transmission is sampled with a uniform $7\times 7$ shifted mesh of \kperp-points in the 2D-BZ perpendicular to the transport direction, corresponding to $10$ and $16$ \kperp-points in the irreducible 2D BZs of the clean nanocontact and of the nanocontact with CO, respectively.\footnote{The importance of accurately sampling the ballistic transmission of model nanocontacts with extended leads has been discussed by \citeauthor{thygesen2005}.\cite{thygesen2005}
We checked that our sampling gives well converged transmission values in the neighborhood of the Fermi level \ef\ (errors within $1\%$), but can result in larger errors at some scattering energies further away from \ef.
However, we are here more interested in the nanocontact conductance and hence we need a well converged transmission just close to \ef.}
Smoother junctions have been simulated using periodic cells with a $(3\times 3)$ in-plane periodicity and pyramidal tips connecting the chain apexes to the Au(001) surfaces (see \pref{fig:autipstran}).
This geometry is composed of a seven-layer slab with an interlayer spacing equal to \db, plus four additional Au atoms in the positions of an additional layer at a distance \ds\ from the surface planes, with the chain attached to the 4-fold hollow sites formed by these additional atoms.
A $5\times5$ uniform mesh of \kperp-points has been used to sample the transmission in the 2D-BZ.

\section{Geometry and energetics}\label{sec:geom}

For different values of the inter-surface distance \dsurf\ (cf.\ \pref{fig:geom}), we consider straight monatomic chains without CO and chains with CO adsorbed at the bridge or atop sites.
In this way, by increasing \dsurf\ we mimic the increase of strain produced on the chain by the pulling of the contact ends.
We optimize the atomic positions of the C and O atoms, and of the Au atoms belonging to the chain, while the positions of all Au atoms in the (001) planes are kept fixed (as are the distances between the planes).
Since we are mainly interested in the local interaction between CO and the Au chain, which is primarily affected by the chain strain, we will not consider here the atomic relaxations of the Au planes at the two sides of the junction.
The effects of these relaxations on the strain dependence of the ballistic conductance are studied in \pref{sec:disc} through an approximate model of the elastic response of the atomic planes to the contact stretching.

\begin{table}[tb]
\caption{Optimized distances (in \ang) and chemisorption energies (in \ev) of CO at the bridge site of a 4-atom-long chain (cf.\ \pref{fig:geom}a) obtained within GGA for selected values of \dsurf.
The optimized distances of the clean chain are also reported.
Owing to the symmetry, 
$d_{3-4}=d_{1-2}$ and $d_{\mathrm{4s}}=d_{\mathrm{s1}}$.}\label{tab:ausurfgeombridge}
\begin{tabular}{c@{\hspace{9pt}}ccc@{\hspace{9pt}}cccccc}
  \hline\hline
    & \multicolumn{3}{c@{\hspace{9pt}}}{4-Au chain} & \multicolumn{6}{c}{CO at the bridge site} \\ 
\dsurf& \dsm& \dab& \dbc& \dsm& \dab& \dbc& \dauc& \dco& \echem\\ \hline
11.56& 1.87& 2.61& 2.61& 1.90& 2.59& 2.86& 2.03& 1.17& $-$1.47\\ 
12.16& 1.99& 2.72& 2.74& 1.93& 2.61& 3.09& 2.02& 1.17& $-$1.61\\
12.76& 2.09& 2.82& 2.94& 1.99& 2.67& 3.44& 2.06& 1.18& $-$1.94\\ 
  \hline\hline
\end{tabular}
\end{table}

\subsection{Clean nanocontact geometries}
We first consider a straight chain without CO and optimize the $z$ coordinate of the \nau\ Au atoms of the chain, thus removing the constraint of uniform interatomic spacing that was adopted in the infinite chain model.\cite{sclauzero2012a}
We did not explore here zigzag or bent configurations, which are expected to become favored only at low values of \dsurf\ in the clean chain.\citep{hakkinen00,sanchez99,skorodumova2005}
The optimized distances between the Au atoms in the chain (\dab, \dbc, \dots) and the distance between the surface plane and the apex atom of the chain (\dsm) are presented in the left parts of \pref{tab:ausurfgeombridge} and \pref{tab:ausurfgeomontop} for a 4-atom-long and a 5-atom-long chain, respectively.
We report here only GGA data, since the LDA results give the same qualitative picture.\cite{sclauzero2010}
In the first row of each table, \dsurf\ is chosen to give Au-Au distances in the short chain close to the equilibrium value in the infinite chain ($2.61\;\ang$), while in the second and third rows the selected $d_{\rm ss}$ values result in moderately or highly stretched Au-Au bonds, respectively.

At the lowest strain considered here, the atoms in the chain are almost equally spaced, but the Au-Au bond length at the extremities of the chain, \dab, adjusts to a value slightly smaller than in the middle (only \dbc\ in the 4-atom and 5-atom chains, but we verified that this holds also for the inner bonds of longer chains).\cite{sclauzero2010}
As the nanocontact is stretched, the distance \dab\ becomes progressively shorter than \dbc.
By studying longer chains ($\nau=6$ and $\nau=7$, not reported here), we observe Au-Au bond lengths which increase while going from the ends toward the center of the unstrained chains. 
Instead, when the average Au-Au bond length is above $3.0\,\ang$ the chains show a tendency to dimerization, with alternating longer and shorter bonds, as already reported in the literature.\cite{okamoto1999}

\subsection{Geometries with adsorbed CO}

We now consider the nanocontact geometries with an adsorbed CO molecule.
The atomic structure is partially optimized as described above and the positions of the Au atoms in the chain are fully relaxed. 
We first consider the bridge geometry, where the CO is placed upright at the bridge site between the two central Au atoms ($\mathrm{Au}_{(2)}$ and $\mathrm{Au}_{(3)}$), as shown in \pref{fig:geom}a.
In \pref{tab:ausurfgeombridge} (right side), we report the optimized distances for the three \dsurf\ values considered before.
The carbon-oxygen bond length in the adsorbed molecule (\dco) is equal or slightly larger than in the infinite chain model\cite{sclauzero2012a} (by less than $1\%$), while C-Au bond lengths (\dauc) are $1\%$ to $3\%$ larger.
We also notice that the bond length between the two Au atoms in contact with CO (\dbc) is always longer than the other Au-Au bonds in the chain ($d_{1-2} = d_{3-4}$).
This Au-Au bond softening in correspondence of the adsorption site is observed also for longer chains.\cite{sclauzero2010}
At the lowest strain studied here ($\dsurf=11.56\,\ang$), \dbc\ adjusts to a value similar to or slightly larger than the Au-Au distance corresponding to the bridge energy minimum for the infinite chain geometry\cite{sclauzero2012a} at $\dauau=2.87\,\ang$.
The other Au-Au bond lenghts, instead, are closer to the equilibrium spacing of the infinite chain.
At low and moderate strains, the chain in its relaxed geometry bends towards CO, while at larger strains the chain atoms get progressively more aligned forming an almost linear strand with one overstretched Au-Au bond in correspondence of the adsorption site.
Because of the flexibility of the Au-C-Au bond angle, very large Au-Au distances between the two atoms in contact with CO are possible as \dsurf\ is increased, but the tilting of the molecule axis from the perpendicular position could become favorable above a critical value of \dsurf.\cite{strange2006}
However, the upright bridge position in the infinite chain geometry\cite{sclauzero2012a} is the lowest energy configuration for Au-Au distances up to $4.2\,\ang$, a distance longer than the largest $d_{2-3}$ reported in \pref{tab:ausurfgeombridge}.
Hence, one can expect that the perpendicular position of CO is preferred for all values of \dsurf\ considered here.

\begin{table}[tb]
\caption{Optimized distances (in \ang) and chemisorption energies (in \ev) of CO at the atop site of a 5-atom long chain (\pref{fig:geom}b). Here, $d_{4-5}=d_{1-2}$, $d_{3-4}=d_{2-3}$, and $d_{\mathrm{5s}}=d_{\mathrm{s1}}$.}\label{tab:ausurfgeomontop}
\begin{tabular}{c@{\hspace{9pt}}ccc@{\hspace{9pt}}cccccc}
  \hline\hline
    & \multicolumn{3}{c@{\hspace{9pt}}}{5-Au chain} & \multicolumn{6}{c}{CO at the atop site} \\ 
\dsurf& \dsm& \dab& \dbc& \dsm& \dab& \dbc& \dauc& \dco& \echem\\ \hline
14.16& 1.86& 2.61& 2.61& 1.98& 2.72& 2.76& 1.98& 1.14& $-$0.78\\ 
14.96& 2.00& 2.73& 2.75& 1.99& 2.73& 2.83& 1.99& 1.14& $-$0.55\\
15.76& 2.09& 2.87& 2.92& 2.04& 2.77& 3.08& 1.98& 1.14& $-$0.63\\ 
  \hline\hline
\end{tabular}
\end{table}

In \pref{tab:ausurfgeomontop}, we report the optimized distances for CO adsorbed atop the central atom of a five-atom-long Au chain ($\mathrm{Au}_{(3)}$ in \pref{fig:geom}b). 
The carbon-oxygen bond distance \dco\ changes very little with strain and is very close to the value found in the infinite chain model.\cite{sclauzero2012a}
Also the C-$\mathrm{Au}_{(3)}$ bond length is not much influenced by strain and is only about $1\%$ larger than the corresponding distance for CO on the infinite chain.
As can be seen from the two nanocontact geometries reported in \pref{fig:geom}, the adsorption of CO in the atop position gives rise to larger distortions of the Au chain compared to the bridge position, with the atom which binds to C, $\mathrm{Au}_{(3)}$, moving towards the molecule and the two lateral atoms, $\mathrm{Au}_{(2)}$ and $\mathrm{Au}_{(4)}$, slightly displaced downwards to create a zigzag geometry.
By comparing \dab\ with \dbc, we see that the binding between C and $\mathrm{Au}_{(3)}$ weakens the metallic bond between $\mathrm{Au}_{(3)}$ and the two neighbouring atoms, $\mathrm{Au}_{(2)}$ and $\mathrm{Au}_{(4)}$.
Indeed, when CO is adsorbed atop the ratio $\dbc/\dab$ is always larger than in the pristine 5-atom-long chain and grows more rapidly with strain (cf.\ \pref{tab:ausurfgeomontop}).

\subsection{Chemisorption energies}
We now turn to examine the chemisorption energies of CO as a function of the surface-surface distance $\echem(\dsurf)$, which are reported in \pref{tab:ausurfgeombridge} and \pref{tab:ausurfgeomontop} for the bridge and atop geometries, respectively.
The large difference between the chemisorption energies of the bridge and atop adsorption sites indicates a strong preference for the former, as already observed in the infinite chain model.\cite{sclauzero2012a}
In the bridge geometry, the chemisorption energy decreases with strain, following the same trend seen for the infinite chain model,\cite{sclauzero2012a} while in the atop geometry this happens only at large enough values of \dsurf.

\begin{figure}[tb]
  \includegraphics[angle=-90,width=0.9\columnwidth]{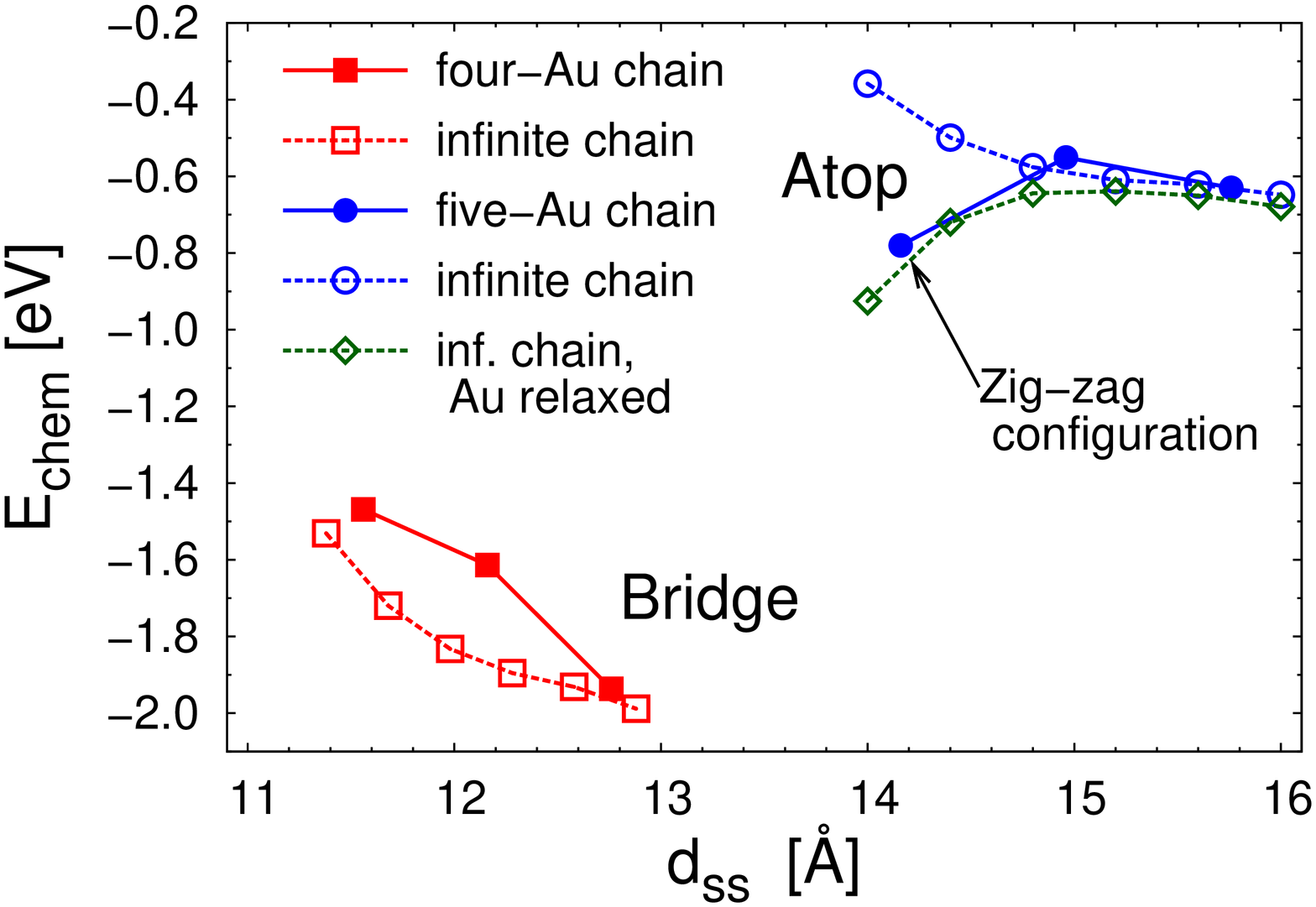}
  \caption{(Color online) Chemisorption energies $\echem(\dsurf)$ for CO at the bridge site of a four-Au-atom chain (filled squares) and at the atop site of a five-Au-atom chain (filled circles).
  The values of $\echem(\dwire)$ for the infinite straight chain\cite{sclauzero2012a} are reported here as a function of $\tilde{\dsurf}(\dwire)=2\cdot\langle\dsm\rangle + (\nau-1)\cdot\dwire$, where $\nau=4$ for the bridge geometry (open squares) and $\nau=5$ for the atop geometry (open circles). 
  $\langle\dsm\rangle$ is chosen as described in the text.
 For the atop site of the infinite chain we also report $\echem(\tilde{\dsurf})$ after optimizing the position along $x$ of the Au atom below CO (open diamonds).}
 \label{fig:ausurfechem}
\end{figure}

This can be seen more clearly in \pref{fig:ausurfechem}, where we compare the chemisorption energies computed with the nanocontact geometry as a function of \dsurf, $\echem(\dsurf)$, with those obtained from the infinite chain geometry as a function of the uniform spacing of the chain \dwire, $\echem(\dwire)$.\cite{sclauzero2012a} 
To do this, we express $\echem(\dwire)$ as a function of an equivalent inter-surface distance $\tilde{\dsurf}(\dwire)$, which can be directly compared to \dsurf.
Hence, we define: $\tilde{\dsurf}=2\cdot\langle\dsm\rangle + (\nau-1)\cdot\dwire$, where the Au-Au spacing \dwire\ is multiplied by the number of Au-Au bonds in the \nau-atom-long suspended chain and $\langle\dsm\rangle$ accounts for the distance between the surface plane and the apex atom of the chain.
We obtain $\langle\dsm\rangle$ by averaging the \dsm\ values reported in \pref{tab:ausurfgeombridge} for the bridge geometry and in \pref{tab:ausurfgeomontop} for the atop geometry, which give $\langle\dsm\rangle=1.94\,\ang$ and $\langle\dsm\rangle=2.00\,\ang$, respectively.

For the bridge geometry, there is a qualitative agreement between $\echem(\tilde{\dsurf})$ from the infinite straight chain (open squares in \pref{fig:ausurfechem}) and $\echem(\dsurf)$ from the nanocontact geometry with $\nau=4$ (filled squares).
This difference in the numerical values could be mainly imputed to the nonuniform Au-Au spacing, to the bending, or to the short length of the chain in the nanocontact geometry.
For the atop geometry (open and filled circles), there is a larger deviation at low \dsurf\ because of the appearance of zigzag configurations in the nanocontact. 
When the height of the Au atom below CO is optimized also in the infinite chain (open diamonds), the agreement with the chemisorption energy obtained in the nanocontact improves considerably, because that Au atom shows a pronounced displacement toward the CO at low strains.
Consequently, energy contributions to \echem\ due to distortions of the chain further away from the adsorption site should be of smaller importance.

Another point that influences the numerical values of \echem\ is the choice of the clean nanocontact geometry which gives the reference energy entering into the calculation of \echem.
Zigzag and bent configurations have been predicted for short monatomic chains between tips,\citep{hakkinen00} while these have not been considered in the present work.
However, we expect that only the \echem\ values for the smallest \dsurf\ value considered here would be affected if using a nonlinear chain geometry as reference. 
Indeed, zigzag or bent geometries are preferred to the linear chain only for sufficiently small values of  \dsurf, while the linear chain is favored with respect to both zigzag and bent configurations when the average spacing between atoms in the chain is equal or larger than about $2.65\,\ang$ and the three configurations are nearly degenerate for slightly smaller spacings.\cite{hakkinen00} 

Finally, this comparison between the infinite chain and the nanocontact geometry shows that a very simplified model such as the straight infinite chain with uniform spacing gives chemisorption energies of CO in qualitative agreement with those obtained through a more realistic and computationally expensive model, leading to the same site preference prediction.
A more quantitative agreement in \echem\ can be obtained by including a few additional degrees of freedom in the geometry optimizations of the infinite chain, such as the Au atom displacement when CO is atop, or a larger spacing between the two Au atoms in contact with CO in the bridge position, which could be inferred from the residual forces after a partial structural optimization.

\begin{figure*}[tp]
		\includegraphics[width=0.8\textwidth]{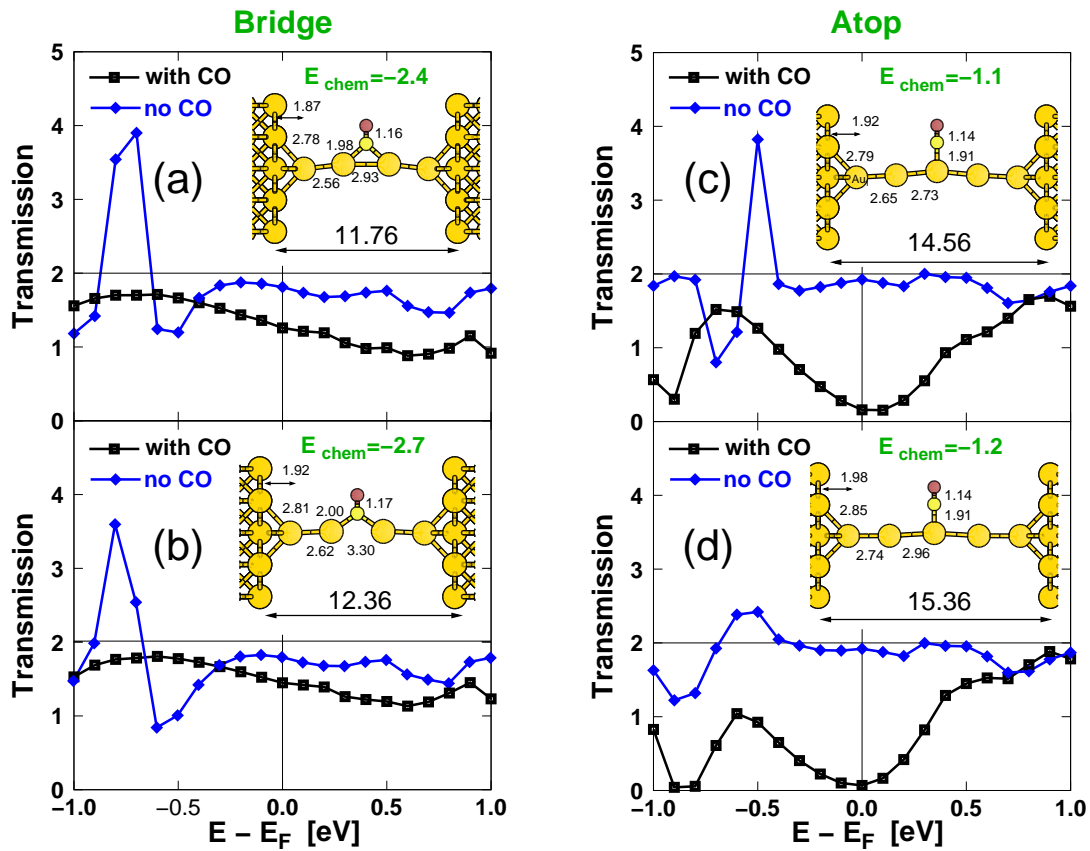}
    \caption{(Color online) Transmission as a function of the electron scattering energy for the clean Au nanocontact (diamonds) and for the Au nanocontact with CO (squares).
    On the left (right) panels, we report the transmission of the clean Au chain with $\nau=4$ ($\nau=5$) and of the chain with CO adsorbed at the bridge (atop) site.
    For each system, two values of the distance between the Au(001) surfaces are considered, a smaller one, corresponding to a moderately strained chain (top panels), and a larger one imposing a high strain on the chain (bottom panels). 
    In each plot, the Fermi energy is indicated by a solid vertical line and the central part of the scattering region with the impurity is shown in the insets.
    The inter-surface distance \dsurf\ and the optimized distances (in \ang), as well as the chemisorption energies of CO (in \ev) are also indicated.
} 
  \label{fig:ausurftran}
\end{figure*}

\begin{figure*}[t]
    \includegraphics[width=0.8\textwidth]{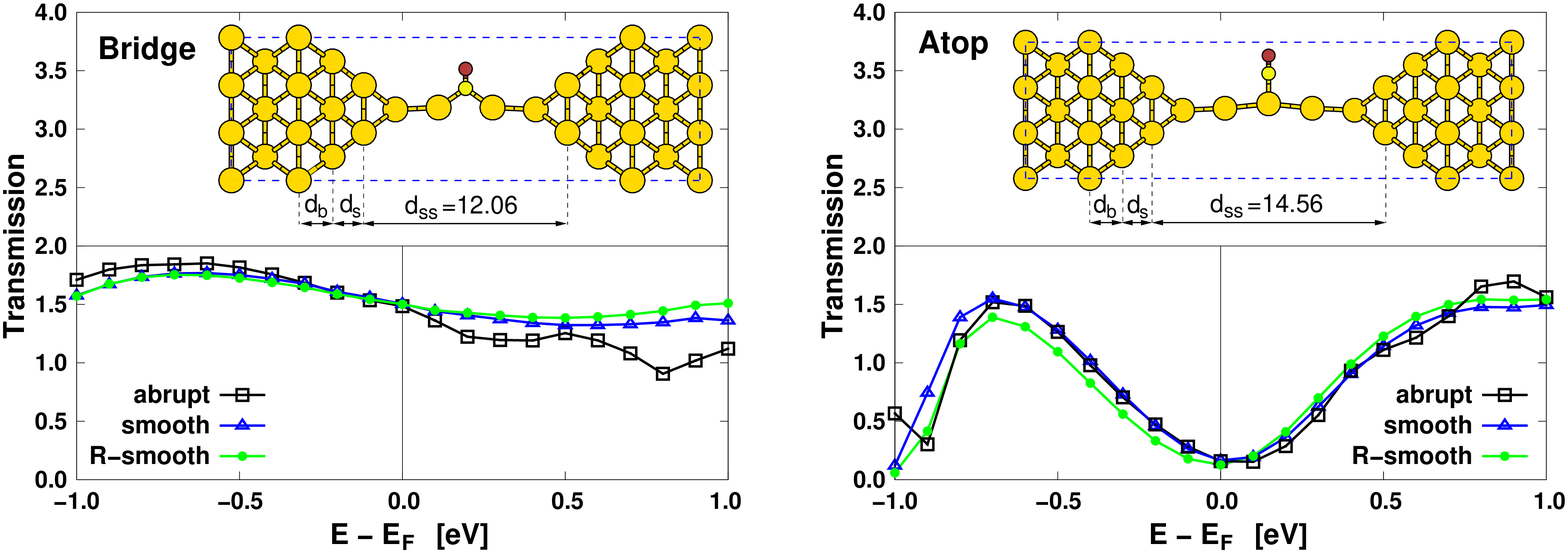}
    \caption{(Color online) Transmission of a short chain with CO adsorbed at the bridge (left) or at the atop site (right): dependence on the smoothness of the surface/chain interface.
     The transmission of the $(2\sqrt{2}\times2\sqrt{2})\mathrm{R}45^{\circ}$ cell (abrupt interfaces, see \pref{fig:ausurftran}) and that of the $(3\times3)$ cell (smoother interfaces, see insets) are compared for a selected value of \dsurf\ and are reported as a function of the scattering energy (squares and triangles, respectively).
    The central region of the smooth junction is built from the atomic positions optimized for the smaller cell (triangles) or by relaxing again the atoms in the constriction (circles).}
  \label{fig:autipstran}
\end{figure*}

\section{Ballistic conductance}\label{sec:cond}

In this section, we address the effects of CO adsorption on the electron transport of the Au nanocontact for several values of the inter-surface distance, \dsurf.
In this way, we aim to describe the strain dependence of the CO-induced changes in the ballistic transmission and to simulate the evolution of the conductance as the contact gets gradually stretched. 

\subsection{Clean nanocontact}
We start discussing the ballistic transport of the short chain for the abrupt junction without CO.
The ballistic transmission as a function of the electron scattering energy is presented here for the LDA case only.
However, the GGA ballistic conductances reported later in \pref{sec:disc} (\pref{fig:autranexp}) show that the same conclusions could be reached within GGA.
In Figs.~\ref{fig:ausurftran}(a--b), we show the transmission of a 4Au-atom straight chain as a function of the scattering energy for two selected values of the intersurface distance \dsurf, one resulting in a moderately strained chain [$\dsurf=11.76\,\ang$, \pref{fig:ausurftran}(a)], and the other in a highly strained chain [$\dsurf=12.36\,\ang$, \pref{fig:ausurftran}(b)].\footnote{%
A more complete discussion of the LDA geometries can be found in Ref.~\onlinecite{sclauzero2010}.
For the readers' convenience, we report here the LDA optimized distances of the clean nanocontact geometries used for the transmission calculations. 
For the 4-Au chain, with $\dsurf=11.76\,\ang$ we obtained $\dsm=1.91\,\ang$, $\dab=2.64\,\ang$, and $\dbc=2.66\,\ang$, while with $\dsurf=12.36\,\ang$ we obtained $\dsm=2.00\,\ang$, $\dab=2.77\,\ang$, and $\dbc=2.81\,\ang$.
For the 5-Au chain, with $\dsurf=14.56\,\ang$ we obtained $\dsm=1.93\,\ang$, $\dab=2.66\,\ang$, and $\dbc=2.69\,\ang$, while with $\dsurf=15.36\,\ang$ we obtained $\dsm=2.01\,\ang$, $\dab=2.81\,\ang$, and $\dbc=2.86\,\ang$.}

In the low strain configuration [\pref{fig:ausurftran}(a)], the conductance is slightly above $0.9\,G_0$, in fair agreement with the theoretical conductance of a 4-atom-chain between Au tips oriented along the [110] direction.\cite{hakkinen00}
Although the number of scattering channels increases with the cross-section of the supercell, for a long enough chain and a small charge transfer between the leads and the chain, the theoretical maximum of the transmission is given by the number of channels in the infinite tipless chain.
Thus, the conductance value close to $1\,G_0$ can be attributed to a single well transmitted spin-degenerate channel of $s$ character.
Around the Fermi energy (\ef) the transmission curve is rather flat, while below \ef\ it is more structured because of a slightly higher reflection of the $s$ channel and the additional contribution from the poorly transmitted $d$ channels of the chain.
This is in line with the common understanding that $d$-type scattering states are more reflected by the presence of an abrupt change in the atomic geometry, like that at the surface/wire interface, owing to the highly directional character of the $d$ wavefunctions. 

When a larger strain is considered [\pref{fig:ausurftran}(b)], the conductance changes very little and decreases of just about $1\,\%$.
Previous calculations\cite{hakkinen00,okamoto1999} have also found that the conductance decreases monotonically when the junction gets stretched and that the conductance has not yet started to drop significantly for average Au-Au spacings around $2.80\,\ang$, being still close to $1\,G_0$,\citep{okamoto1999} or slightly below.\citep{hakkinen00}
Above \ef\ the shape of the transmission function is very similar to that obtained for the low-strain configuration, while below \ef\ the reflection slightly increases with strain. 

In Figs.~\ref{fig:ausurftran}(c--d), we show the transmission function for the 5-atom-long chain for $\dsurf=14.56\,\ang$ and for $\dsurf=15.36\,\ang$, corresponding to moderate and high strains, respectively.
The conductance of the 5Au-atom chain is about $0.96\,G_0$ for both \dsurf\ values considered here and therefore is slightly larger than that of the 4Au-atom chain, in agreement with the odd-even effect seen in experiments\citep{smit2003} and in theoretical calculations.\citep{delavega2004,skorodumova2005}
For instance, \citet{delavega2004} have calculated the theoretical conductance of short chains between flat Au(111) surfaces showing that the conductance for an even number of atoms in the chain is lower than that for an odd number of atoms, as also seen in experiments.
Their conductance values, about $0.96\,G_0$ and $0.99\,G_0$ for 4-atom and 5-atom long chains, respectively, are somewhat larger compared to the results of this work, probably because their chains are attached to the more compact Au(111) surface.
Also for the 5Au-atom chain, we find that the conductance dependence on strain is quite modest, since it stays almost constant when the average spacing in the chain grows from about $2.74\,\ang$ to $2.90\,\ang$ as $\dsurf$ is increased.
For the 5-atom-long chain between Au(111) surfaces, de la Vega and coworkers have found that the conductance goes from $0.99\,G_0$ to about $1\,G_0$ when the spacing between atoms in the chain grows from $2.70\,\ang$ to $3.00\,\ang$, while the conductance of the 4-atom-long chain does not change appreciably,\citep{delavega2004} as in the case of the Au(001) surface examined here.

\subsection{Nanocontact with adsorbed CO}
We will now describe the changes in the ballistic transmission induced by CO adsorption at the bridge site or at the atop site of the Au chain in the nanocontact.
The transmissions are shown as a function of the scattering energy in Figs.~\ref{fig:ausurftran}(a--b) for the bridge geometry and in Figs.~\ref{fig:ausurftran}(c--d) for the atop one (squares), together with the previously discussed transmissions of the clean nanocontact (diamonds) for the same values of \dsurf. 
 
When CO is at the bridge site, the conductance of the chain at low strains [\pref{fig:ausurftran}(a)] is reduced to about $0.63\,G_0$, a value close to the fractional conductance peak seen in Au nanocontacts experiments in presence of CO gas.\citep{kiguchi2007}
The transmission curve shows only a slight dependence on the scattering energy and decreases monotonically in the range of energies considered here.
The $5d$ transmission peak seen below \ef\ in the pristine chain is suppressed here, but actually for energies slightly above or below it the transmission increases after CO adsorption.
The dependence of the transmission on the scattering energy around \ef\ is similar to that seen in Ref.~\onlinecite{sclauzero2012a} for the tipless infinite chain geometries at medium/high strains, but the transmission values are lower by about $10\,\%$ in the short chain geometries.
At larger strains [\pref{fig:ausurftran}(b)], the transmission of the bridge configuration is still characterized by a smoothly varying and monotonically decreasing behaviour as a function of energy, but shows slightly higher values with respect to the low strain configuration.
The conductance grows to about $0.72\,G_0$ and is therefore compatible with the conductance increase due to the contact stretching observed in the experimental Au conductance traces at fractional conductance values.\citep{kiguchi2007} 
Previous conductance calculations for a 3-atom chain between Au(111) leads and a CO molecule at the bridge have found a larger conductance,\cite{xu2006} about $0.9\,G_0$, for a strain level which is roughly intermediate between the two considered here.
This discrepancy may be ascribed to the shorter chain length\citep{grigoriev2006} or to the rather different functional used (B3LYP).

When CO is adsorbed at the atop site, the conductance reduction with respect to the clean nanocontact is much stronger than for the bridge site.
In the low-strain configuration [\pref{fig:ausurftran}(c)], the transmission curve has a wide depression centered just above \ef\ and the conductance is about $0.08\,G_0$, more than one order of magnitude smaller than in the clean nanocontact.
Below \ef, the $d$ peak is suppressed as in the bridge geometry, while at energies higher than $0.5\,\ev$ the $s$ channel is less reflected and has a transmission which approaches that of the pristine chain.
At higher strains [\pref{fig:ausurftran}(d)], the transmission dip shifts towards \ef\ causing a further lowering of the conductance down to about $0.03\,G_0$. 
Therefore, the strain dependence of the conductance in the bridge and in the atop configurations are completely different, as already inferred from the transmission of the infinite chain geometries.\cite{sclauzero2012a}
Moreover, the mechanism suppressing the transmission around the Fermi energy in the atop geometry is the same for both nanocontact and infinite chain geometries. 
It can be regarded as a result of Fano-like destructive interference due to resonance scattering on the \sga\ antibonding state brought up by CO adsorption and appearing right at \ef\ in the nanocontact geometry.
A more complete picture of the connection between the electronic structure features and the ballistic transmission of the CO/Au chain system was presented in a previous work using a tipless infinite chain model.\cite{sclauzero2012a}

These conclusions are not modified when considering ``smoother'' junctions with pyramidal tips on both sides of the chain.
We use the previously relaxed ``abrupt'' geometries to build a smoother junction by inserting the CO and the Au chain between the two terminal atomic planes of the tips (see insets of \pref{fig:autipstran}).
These planes are made by four atoms each, arranged to form a square, and are spaced by \dsurf\ along $z$.
In \pref{fig:autipstran}, we compare the transmission curves of the so-obtained ``smooth'' interface (triangles) with those of the corresponding ``abrupt'' interface (squares) for a selected value of \dsurf, both for the bridge and for the atop geometry.
A quite good agreement can generally be observed below \ef, while above \ef\ there are larger discrepancies, especially for the bridge geometry. 
Nevertheless, the almost perfect correspondence of the conductance values confirms that the striking difference between the bridge and atop conductances is not influenced by the precise atomic geometry chosen to model the wire/surface interface.
We further optimized part of the structure by letting the 4 basal atoms on each side move along the longitudinal direction ($z$ axis) and completely relaxing the atomic positions of the chain and of the CO.
The structural changes are rather small and the total energy is lowered by only $0.3\,\ev$ or less.
The transmissions obtained from these relaxed smooth interfaces (dots in \pref{fig:autipstran}) do not differ appreciably from the non-relaxed structure in the bridge geometry, while the \sga\ transmission dip moves slightly closer to \ef\ in the atop geometry.
In both cases, these conductance values are very similar to those presented above and do not affect the conclusions drawn from the analysis of the abrupt interfaces.

\section{Discussion and comparison with experimental data}\label{sec:disc}

A reproducible behaviour in the conductance traces of Au nanocontacts in presence of CO was reported by a recent experiment:\citep{kiguchi2007} a sharp reduction of the conductance, from the $1\,G_0$ value of the Au monatomic chain down to about $0.5\,G_0$, followed by a slow increase upon further stretching of the nanocontact and by the final drop into the tunneling regime after a small elongation. 
It is reasonable to assume that, at some point of the pulling cycle, a CO molecule sticks to the Au chain giving rise to the sudden reduction of the conductance.
Our theoretical conductances for the bridge and atop geometries, together with their strain dependence, allow us to simulate the conductance trace after CO adsorption at the bridge or at the atop site, and hence to make a direct comparison with the experimental traces.

\begin{figure}[tb]
    \includegraphics[width=0.9\columnwidth]{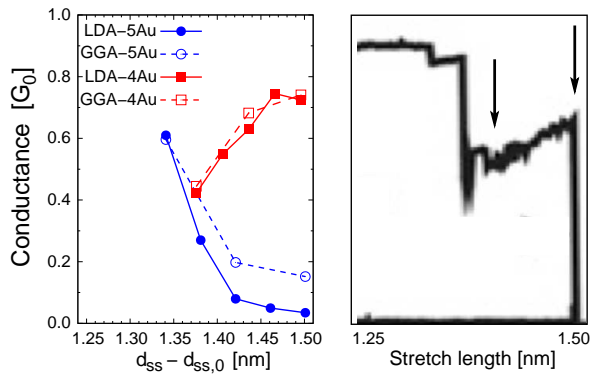}
    \caption{(Color online) Ballistic conductance as a function of the tips displacement: comparison between simulations (left panel) and experiment (right panel).
    Both LDA (open symbols) and GGA (filled symbols) conductances are shown as a function of \dsurf\ for the bridge (squares) and atop (circles) geometries. 
    \dsurf\ is shifted by a constant value $d_{\mathrm{ss},0}$ to keep into account the experimental offset in the displacement measurement.
    The experimental conductance trace has been adapted from Ref.~\onlinecite{kiguchi2007}.
    The same scales for the conductance and the displacement are used in the two panels.}
  \label{fig:autranexp}
\end{figure}

In \pref{fig:autranexp}, the conductances of the bridge and atop geometries are shown as a function of the displacements between the Au surfaces and are juxtaposed with an experimental conductance trace\cite{kiguchi2007} which presents the features described here above. 
We notice that the bridge configuration gives conductance values close to experimental ones and the correct dependence on strain, while the atop one gives the opposite behaviour with strain and too low conductance values at high strains. 
In the bridge geometry, the GGA and LDA conductances show a very good agreement with each other.
In the atop geometry, the GGA conductance decreases more slowly than the LDA one, but the dependence on the contact stretching is the same.
These conductance results, together with the strong energetic preference for the bridge site (\pref{sec:geom}), are compatible with the presence of the fractional peak at about $0.6\,G_0$ in the Au conductance histograms and with the absence of a low conductance tail, which would be possible only in presence of an energetically favored atop geometry.

However, we notice that the theoretical conductance of the bridge geometry reproduces the slope of the experimental conductance trace only qualitatively (\pref{fig:autranexp}).
We are not aware of other experiments reporting about the conductance slope of the CO/Au nanocontact system and we are not able to assess the degree of the experimental reproducibility, but we find anyway interesting to explore further this point.
The discrepancy in the numerical values of the slope might be related to the elastic response of the bulk leads to the external pulling force, which is present in the real nanocontact but has been neglected in the atomic relaxations of our model geometries. 
Indeed, a more realistic modeling of the structural modifications during the contact stretching process would require the relaxation of the atomic position in the atomic planes forming the leads, but this can be done only considering much larger supercells.

It is possible to keep into account approximately the mechanical response of the leads to the external stress by treating them as ideal springs with a finite spring-constant $k_s$.\cite{strange2006}
We assume that the position of the electrodes is controlled at two opposite ends far away from the junction, separated by a distance $L \gg \dsurf$.
The force balance between the ideal springs and the elastic response of the junction region can be expressed by the following equation:\citep{strange2006}
\begin{equation}
  \frac{1}{2} k_s (L - \dsurf) = \frac{\de \etot(\dsurf)}{\de \dsurf},
  \label{eq:springs}
\end{equation}
where $\etot(\dsurf)$ is the total energy of the relaxed nanocontact geometry with inter-surface distance \dsurf.
By solving for $L$ in \pref{eq:springs}, we can convert the inter-surface distance \dsurf\ into an equivalent tips displacement $L(\dsurf;k_s$), which includes the elongation of the leads and has only the electrode stiffness $k_s$ as external parameter.
Notice that in the unrealistic assumption of infinitely stiff leads ($k_s=\infty$), the external stress is totally released in the nanocontact region that has been atomically relaxed (the short monatomic chain) and we have $L=\dsurf$.
Experimental estimates of $k_s$ in Au electrodes have been reported on the basis of the non-exponential dependence of the tunneling current on the distance: values in the range $0.3\,\ev/\ang^2$ to $3.7\,\ev/\ang^2$ have been found, depending on the junction realization.\cite{rubio-bollinger2004} 

\begin{figure}[tb]
    \includegraphics[width=0.8\columnwidth]{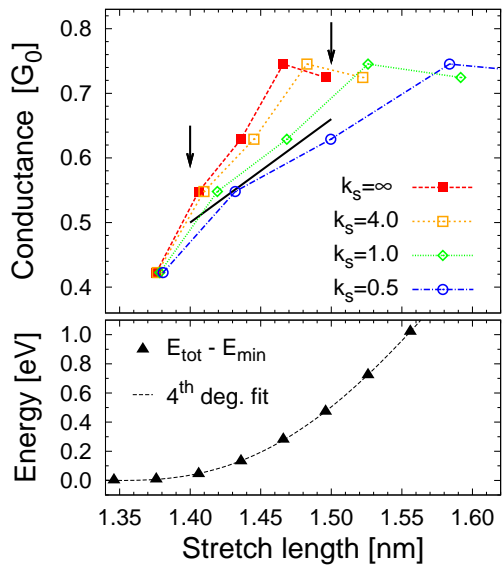}
    \caption{(Color online) Top panel: ballistic conductance of the 4Au-atom nanocontact with CO at the bridge site as a function of the equivalent tips displacement $L$ for different values of the electrode stiffness $k_s$ (in $\ev/\ang^2$).
    The arrows point the start and end points indicated by the arrows in \pref{fig:autranexp} and the solid line between those points approximates the experimental conductance trace\cite{kiguchi2007} with a linearly increasing conductance.
    Bottom panel: total energy of the optimized nanocontact geometry as a function of the inter-surface distance \dsurf.} 
  \label{fig:autranexp2}
\end{figure}

In \pref{fig:autranexp2} (bottom panel), we report the total energy $\etot(\dsurf)$ of the relaxed LDA bridge geometry for some \dsurf\ values in a range of interest and we fit this energy with a $4^{\rm th}$ degree polynomial.
The analytical derivative of the energy fit is used to obtain $L(\dsurf;k_s$) for a few values of the parameter $k_s$ and then express the \dsurf-dependent conductance as a function of $L$ (see \pref{fig:autranexp2}, upper panel). 
The case of infinitely stiff leads ($k_s=\infty$) corresponds to the conductance already shown in \pref{fig:autranexp} and is reported here again for comparison.
When taking a finite $k_s$, a fraction of the term depending on the energy derivative is added to \dsurf\ in $L$, thus the calculated conductance points are shifted to higher displacements and the theoretical conductance slope decreases.
For $k_s$ values close to the experimental upper limit ($4.0\,\ev/\ang^2$) the slope does not change significantly, but for $k_s$ values between $0.5\,\ev/\ang^2$ and $1.0\,\ev/\ang^2$ (a realistic range, according to the experiment\cite{rubio-bollinger2004}) the calculated conductance slope is in much better agreement with that of the experimental conductance trace\cite{kiguchi2007} (see thick solid line in \pref{fig:autranexp2}). 
This demonstrates that a more accurate description of the leads is needed to obtain a correct slope of the conductance as a function of the contact stretch, but already good estimates can be obtained through this simple elastic-response model.

\section{Conclusions}\label{sec:concl}

We have studied through density functional calculations the adsorption of CO on Au monatomic chains in a model nanocontact geometry and its effects on the ballistic conductance of the nanocontact as a function of the contact stretching.
By comparing the adsorption energies of CO at the bridge and atop sites, we find that the bridge site is energetically favored at all levels of Au strain, as also found in a simpler model without bulk leads.\cite{sclauzero2012a}
The chemisorption energies of CO on the short chain in the nanocontact geometry are comparable to those found in the infinite straight chains\cite{sclauzero2012a} and a fair agreement in the variation of the energetics with strain is obtained when the displacement of the Au atom below the molecule in the infinite chain geometry is taken into account.
This was not clearly predictable, because the finite length of the chain and the non-uniform Au-Au spacing are not encompassed by the infinite chain model, while in the nanocontact geometries these aspects are included and play a role in the structural modifications of the short chain after CO adsorption.
For instance, when CO is at the bridge site, at low strains the chain bends towards the molecule, while at larger strains the Au-Au bond below CO elongates much more rapidly than the others.
With CO atop, the chain forms a zigzag geometry, more pronounced at lower strains, where the Au atom right below the CO moves towards the molecule.

The electron transmission across the Au nanocontact in presence of CO displays some important features which have already been found in the tipless geometries, most notably the transmission dip close to the Fermi level in the atop geometry which causes a strong suppression of the conductance.
This dip is not present in the bridge geometry at those scattering energies, therefore the conductance reduction with respect to the pristine nanocontact is much smaller. 
Also the dependence of the ballistic conductance on strain does not change with the inclusion of the tips, modeled here using either abrupt interfaces between the chain and the Au(001) surfaces or smoother pyramidal junctions: the bridge conductance is found to increase slightly with strain, compatible with the experimental findings, while the atop conductance drops rapidly down to zero as the contact is stretched because of the \sga\ transmission dip moving closer to \ef.
With the inclusion of tips, the bridge geometry shows conductance values close to those of the experimentally observed structure which forms in the Au nanocontact after CO exposition.
The slope of the experimental conductance with respect to the contact stretching can also be reproduced with reasonable accuracy by our calculations if the tips displacement is computed taking into account the elastic response of the bulk leads through a simple model.

\begin{acknowledgments}
The authors are grateful to Erio Tosatti for useful and stimulating discussions.
This work has been supported by PRIN-COFIN 20087NX9Y7, as well as by INFM/CNR ``Iniziativa trasversale calcolo parallelo''. 
All calculations have been performed on the SISSA-Linux cluster and at CINECA in Bologna.
\end{acknowledgments}


%

\end{document}